\documentclass[prd,preprint,nofootinbib]{revtex4-1} 

\usepackage{latexsym,verbatim}
\usepackage{amssymb}
\usepackage{amsfonts}
\usepackage{amsmath,color}
\usepackage{graphicx} 
\usepackage{bm}
\usepackage[utf8]{inputenc}
\usepackage[T1]{fontenc}

\usepackage{hyperref}

\hypersetup{
    colorlinks=true,
    linkcolor=red,
    citecolor=red,
    filecolor=magenta,      
    urlcolor=cyan,
    pdftitle={Effects of gravitational waves...},
    pdfpagemode=FullScreen,}

\def\mb#1{\mathbf{#1}}

\def\ber{\begin{eqnarray}}
\def\eer{\end{eqnarray}}
\def\beq{\begin{equation}}
\def\eeq{\end{equation}}

\def\rmd{{\rm d}}

\def\ed{\end{document}}

\def\sT{\sin \left(\omega T \right)}
\def\cT{\cos \left(\omega T \right)}

   \let\square=\dal

\def\fp{f^{+}}
\def\fc{f^{\times}}

\makeatletter
\let\jnl@style=\rm
\def\ref@jnl#1{{\jnl@style#1}}

\def\aj{\ref@jnl{AJ}}                   
\def\actaa{\ref@jnl{Acta Astron.}}      
\def\araa{\ref@jnl{ARA\&A}}             
\def\apj{\ref@jnl{ApJ}}                 
\def\apjl{\ref@jnl{ApJ}}                
\def\apjs{\ref@jnl{ApJS}}               
\def\ao{\ref@jnl{Appl.~Opt.}}           
\def\apss{\ref@jnl{Ap\&SS}}             
\def\aap{\ref@jnl{A\&A}}                
\def\aapr{\ref@jnl{A\&A~Rev.}}          
\def\aaps{\ref@jnl{A\&AS}}              
\def\azh{\ref@jnl{AZh}}                 
\def\baas{\ref@jnl{BAAS}}               
\def\bac{\ref@jnl{Bull. astr. Inst. Czechosl.}}
\def\caa{\ref@jnl{Chinese Astron. Astrophys.}}
\def\cjaa{\ref@jnl{Chinese J. Astron. Astrophys.}}
\def\icarus{\ref@jnl{Icarus}}           
\def\jcap{\ref@jnl{J. Cosmology Astropart. Phys.}}
\def\jrasc{\ref@jnl{JRASC}}             
\def\memras{\ref@jnl{MmRAS}}            
\def\mnras{\ref@jnl{MNRAS}}             
\def\na{\ref@jnl{New A}}                
\def\nar{\ref@jnl{New A Rev.}}          
\def\pra{\ref@jnl{Phys.~Rev.~A}}        
\def\prb{\ref@jnl{Phys.~Rev.~B}}        
\def\prc{\ref@jnl{Phys.~Rev.~C}}        
\def\prd{\ref@jnl{Phys.~Rev.~D}}        
\def\pre{\ref@jnl{Phys.~Rev.~E}}        
\def\prl{\ref@jnl{Phys.~Rev.~Lett.}}    
\def\pasa{\ref@jnl{PASA}}               
\def\pasp{\ref@jnl{PASP}}               
\def\pasj{\ref@jnl{PASJ}}               
\def\rmxaa{\ref@jnl{Rev. Mexicana Astron. Astrofis.}}%
\def\qjras{\ref@jnl{QJRAS}}             
\def\skytel{\ref@jnl{S\&T}}             
\def\solphys{\ref@jnl{Sol.~Phys.}}      
\def\sovast{\ref@jnl{Soviet~Ast.}}      
\def\ssr{\ref@jnl{Space~Sci.~Rev.}}     
\def\zap{\ref@jnl{ZAp}}                 
\def\nat{\ref@jnl{Nature}}              
\def\iaucirc{\ref@jnl{IAU~Circ.}}       
\def\aplett{\ref@jnl{Astrophys.~Lett.}} 
\def\apspr{\ref@jnl{Astrophys.~Space~Phys.~Res.}}
\def\bain{\ref@jnl{Bull.~Astron.~Inst.~Netherlands}}
\def\fcp{\ref@jnl{Fund.~Cosmic~Phys.}}  
\def\gca{\ref@jnl{Geochim.~Cosmochim.~Acta}}   
\def\grl{\ref@jnl{Geophys.~Res.~Lett.}} 
\def\jcp{\ref@jnl{J.~Chem.~Phys.}}      
\def\jgr{\ref@jnl{J.~Geophys.~Res.}}    
\def\jqsrt{\ref@jnl{J.~Quant.~Spec.~Radiat.~Transf.}}
\def\memsai{\ref@jnl{Mem.~Soc.~Astron.~Italiana}}
\def\nphysa{\ref@jnl{Nucl.~Phys.~A}}   
\def\physrep{\ref@jnl{Phys.~Rep.}}   
\def\physscr{\ref@jnl{Phys.~Scr}}   
\def\planss{\ref@jnl{Planet.~Space~Sci.}}   
\def\procspie{\ref@jnl{Proc.~SPIE}}   

\makeatother

\begin{document}

\author{Matteo Luca Ruggiero}
\email{matteoluca.ruggiero@unito.it}
\affiliation{Dipartimento di Matematica ``G.Peano'', Universit\`a degli studi di Torino, Via Carlo Alberto 10, 10123 Torino, Italy}
\affiliation{INFN - LNL , Viale dell'Universit\`a 2, 35020 Legnaro (PD), Italy}

\date{\today}

\title{Effects of gravitational waves on electromagnetic fields}

\begin{abstract}
{We study the interaction between a plane gravitational wave and electromagnetic fields, analyzing this interaction in the proper detector frame. The gravitational field is treated as an effective electromagnetic medium, and within this framework, we demonstrate that the coupling between pre-existing electromagnetic fields and the gravitational wave generates new effective currents. This approach, an alternative to previously explored methods, has the advantage of employing Fermi coordinates, which enable direct reference to measurable quantities.}
To assess the impact of the effect of gravitational waves, we solve Maxwell's equations for some standard configurations of the electric and magnetic fields.
\end{abstract}

\maketitle

\section{Introduction} \label{sec:intro}

After their first direct interferometric detection  \cite{LIGO2016}, gravitational waves  have opened a new window for research in astrophysics and cosmology: in fact, gravitational waves are important not only because they constitute yet another test of Einstein's theory \cite{2014arXiv1409.7871W}, but also because  they serve as crucial instruments for exploring the universe in the age of multi-messenger astronomy. Advancements in technology and specialized missions are poised to significantly enhance the wealth of information accessible through this avenue (for further details, see, e.g., \citet{2019Natur.568..469M,2021NatRP...3..344B} and related references). 

Furthermore, experiments with gravitational waves are crucial for testing gravitational theories beyond General Relativity. Indeed, challenges to Einstein's theory emerge from observations of the universe on a large scale, with the problems of dark matter and dark energy  \cite{universe2040023}, and at a more fundamental level we still don't know how to reconcile General Relativity with the Standard Model  of particle physics. In order to  try to cope with these issues, theories with additional fields (such as scalar ones) or provided with a richer geometrical structure were proposed and, in these cases, gravitational waves  might have different features with respect to the general relativistic ones, such as for instance longitudinal effects (see e.g. \citet{capozziello2}), or the generation process may have pecularities, such as in the Einstein-Cartan theory \cite{Battista:2021rlh,Battista:2022hmv}; {we remember that the latter can be thought of as an extension  of General Relativity to Riemann-Cartan spaces, where both the metric and the torsion determine the geometry of space-time \cite{Ruggiero:2003tw}. Interestingly, formal parallels were highlighted between the General Relativity treatment of bodies with macroscopic angular momentum and the Einstein-Cartan description of spinning objects modeled using the Weyssenhoff fluid \cite{battista2,battista1}.}

These motivations drive enormous efforts to further develop current interferometers, such as LIGO and VIRGO, as well as to design and create brand new ones, such as LISA.

More generally, investigations on gravitational waves  do not solely rely on interferometers, where changes in relative distances between test masses are measured using lasers. Other Earth-based experimental setups have been proposed, including  accelerators \cite{acceleratori,acceleratori2}, storage rings \cite{storage}, two-level system resonance \cite{Ruggiero_2020b} and microcavities  \cite{Berlin:2021txa}. In the latter case, particularly, researchers study the effects of gravitational waves on electromagnetic fields in resonant cavity experiments. So, the interaction between gravitational waves and electromagnetic fields, which has extensively been studied (see e.g. \citet{Cooperstock:1993nn,Fortini:1996fpz,Montanari:1998gd,Mieling:2021ftj} and references therein) presents potential for exploring new experimental paths.

The purpose of this paper is to develop a formalism to study the effect of a plane gravitational wave on electromagnetic fields. In order to do this, we solve Maxwell's equation in the curved spacetime of a plane gravitational wave and we do this in the proper detector frame. Indeed, gravitational waves are typically studied using the transverse and traceless (TT)  tensor, which allows to introduce the so-called \textit{TT coordinates or TT frame}. The \textit{proper detector frame} is an alternative approach \cite{maggiore2007gravitational} which is based on the use of \textit{Fermi coordinates}. The latter  are a quasi-Cartesian coordinates system that can be build in the neighbourhood of the world-line of an observer, and their definition depends both on the background field where the observer is moving and, also, on the kind of motion. Fermi coordinates are  defined, by construction, as scalar invariants \cite{synge1960relativity}; they have a concrete meaning, since they are the coordinates an observer would naturally use to make space and time measurements in the vicinity of his/her world-line. 

Recently, the interaction of gravitational waves and electromagnetic fields has been studied in the TT frame by \citet{lobo}. {The novelty of this paper lies in the study of the problem within the proper detector frame and in the use of a formalism that allows Maxwell's equations in curved spacetime to be expressed as flat spacetime equations in the presence of an anisotropic medium.} Subsequently, some simplified interaction models will be discussed to understand and evaluate the impact of a plane gravitational wave on the electromagnetic fields already present before its passage.

The paper is organized as follows: in Section \ref{sec:maxs} we introduce the formalism to study Maxwell's equations in curved spacetime, while in Section \ref{sec:fermi} we focus on the construction of the local spacetime metric in the laboratory frame, with the use of Fermi coordinates, and we apply  it to the field of a plane gravitational wave. Next, in Section \ref{sec:modfields}, we present the fundamental equations necessary for investigating the influence of gravitational waves on electromagnetic fields, followed by their solution in select prototypical scenarios. Subsequently, discussion and conclusions are provided in Section \ref{sec:discon}.

\section{Maxwell's equations in curved spacetime} \label{sec:maxs}

We aim to investigate the influence of gravitational waves on electromagnetic fields. To achieve this, we must solve Maxwell's equations in curved spacetime. If $F_{\mu\nu}$ is the antisymmetric tensor of the electromagnetic field, the equations that determine the electromagnetic field in curved spacetime are\footnote{Latin indices refer to space coordinates, while Greek indices to spacetime ones. We will use bold-face symbols like  $\mb W$ to refer to space vectors; the spacetime signature is $+2$ in our convention.} \cite{landau2013classical,MTW}:
\begin{eqnarray}
F_{\mu\nu;\alpha}+F_{\nu\alpha;\mu}+F_{\alpha\mu;\nu} & = & 0, \\ \label{eq:maxc1}
\varepsilon_{0}F^{\mu\nu}_{\,\,\,\,\,;\nu} & = & j^{\mu}, \label{eq:maxc2}
\end{eqnarray}
where ``$;$'' indicates covariant derivative, and $j^{\mu}$ is the four-current. The above equations, due to the antisymmetry of the electromagnetic tensor, can be also written in the form 
\begin{eqnarray}
F_{\mu\nu,\alpha}+F_{\nu\alpha,\mu}+F_{\alpha\mu,\nu} & = & 0, \label{eq:max1} \\
\frac{\varepsilon_{0}}{\sqrt{-g}} \partial_{v}\left(\sqrt{-g} F^{\mu\nu} \right) & = & j^{\mu}, \label{eq:max2}
\end{eqnarray} 
where ``$,$'' stands for partial derivative, and $g$ is the determinant of the spacetime metric. The covariant electromagnetic tensor $F_{\mu\nu}$ is usually defined in terms of the three-dimensional electric and magnetic fields\footnote{We use Cartesian coordinates here and henceforth.} $\mb E$ and $\mb B$

\begin{equation} \label{eq:Fmunu1}
 F_{\mu\nu}=  \begin{pmatrix}
       0 &\  \ E_x & \ \ E_y & \ \ E_z \\
      -E_x & \ \ 0 & \ \ -cB_z & \ \ cB_y \\
      -E_y & \ \ cB_z & \ \ 0 & \ \ -cB_x \\
      -E_z & \ \ -cB_y & \ \ cB_x & \ \ 0  
   \end{pmatrix}.
\end{equation}

{Our approach to solving equations (\ref{eq:max1})-(\ref{eq:max2}) relies on the observation that the influence of a gravitational field on electromagnetic fields can be represented as a material medium. This idea is not new; it traces back to the works  on the propagation of electromagnetic waves in a gravitational field by \citet{optical} and \citet{plebanski}, who noted that the concept of an equivalent material medium was first proposed by \citet{tamm} in 1924. Later, \citet{masshoonprd} utilized this framework to analyze the scattering of electromagnetic radiation by black holes (see also the discussion in \citet{optical0}). More recently, \citet{Leonhardt:2006ai,ulf2} demonstrated the relevance of this approach for designing metamaterials. Specifically,  equations (\ref{eq:max1})-(\ref{eq:max2}) can be simplified and recast in a form resembling Maxwell's equations in flat spacetime within a material medium.} To do this, we introduce  $H^{\mu\nu}$ such that
\beq
H^{\mu\nu}=\varepsilon_0\sqrt{-g}F^{\mu\nu}=
\varepsilon_0\sqrt{-g}g^{\mu\lambda}g^{\nu\rho}F_{\lambda\rho}, \label{eq:Hmunu1}
\eeq
which contains the fields $\mb D$ and $\mb H$ since
\begin{equation} \label{eq:Hmunu2}
 H^{\mu\nu}=  \begin{pmatrix}
       0 &\  \ -D^{x} & \ \ -D^{y} & \ \ -D^{z} \\
      D^{x} & \ \ 0 & \ \ -H^{z}/c & \ \ H^{y}/c \\
      D^{y} & \ \ H^{z}/c & \ \ 0 & \ \ -H^{x}/c \\
      D^{z} & \ \ -H^{y}/c & \ \ H^{x}/c & \ \ 0  
   \end{pmatrix}.
\end{equation}
Now, if we set
\beq
J^{\mu}=\sqrt{-g}j^{\mu}, \label{eq:sources}
\eeq
the source equations (\ref{eq:max2}) can be written in the form
\beq
 \partial_{\nu}H^{\mu\nu}=J^\mu.  \label{eq:Hmunu3}
\eeq
In other words, Maxwell's source equations in curved spacetime are written in the form of electromagnetic equations in flat spacetime in a material medium, with current $J^\mu$, with the constitutive equations given by (\ref{eq:Hmunu1}). The homogenous equations (\ref{eq:max1}) are instead expressed in terms of the fields $\mb E, \mb B$ as in flat spacetime. 
In three dimensional notation we have
\begin{equation}
\quad \nabla\cdot\mathbf{D} = \rho \,,\quad 
\nabla\times\mathbf{H} = \frac{\partial \mathbf{D}}{\partial t}
+ \mathbf{j} \,,\quad
\nabla\cdot\mathbf{B} = 0  \,,\quad 
\nabla\times\mathbf{E} = -\frac{\partial \mathbf{B}}{\partial t}
\,.
\label{eq:maxwell}
\end{equation}

Using Eq. (\ref{eq:Hmunu1}), it is possible to obtain the relation between $\mb E, \mb D, \mb H$. In fact, we get
\beq
D^{j}=-\epsilon_{0} \frac{\sqrt{-g}}{g_{00}}g^{ji}E_{i}-\epsilon_{jik} w_{k} \frac{H_{i}}{c}, \label{eq:const1}
\eeq
where $\displaystyle w_{i}=\frac{g_{0i}}{g_{00}}$; the above equation can be written in three-dimensional form
\beq
\mathbf{D} = \varepsilon_0\varepsilon\,\mathbf{E}
+ \frac{\mathbf{w}}{c}\times\mathbf{H}, \label{eq:const2}
\eeq
whith
\beq
\varepsilon  = -\frac{\sqrt{-g}}{g_{00}}\,g^{ij} .\label{eq:defepsilon}
\eeq
To obtain the other constitutive equations between $\mb B, \mb H, \mb E$, we introduce the dual tensors
$^*\!F^{\mu\nu}$ and $^*\!H_{\mu\nu}$ 
defined by  \cite{Leonhardt:2006ai}
\beq
^*\!F^{\mu\nu}=
\frac{1}{2}e^{\mu\nu\lambda\rho}F_{\lambda\rho} , \label{eq:dual1}
\eeq

\beq
^*\!H_{\mu\nu}= 
\frac{1}{2}e_{\mu\nu\lambda\rho}H^{\lambda\rho},  \label{eq:dual2}
\eeq
where 
\beq
e_{\mu\nu\lambda\rho}=\sqrt{-g}\,  \epsilon_{\mu\nu\lambda\rho}, \quad
e^{\mu\nu\lambda\rho}=
-\frac{1}{\sqrt{-g}}\epsilon_{\mu\nu\lambda\rho}. \label{eq:levicivita}
\eeq
In the above equations $\epsilon_{\mu\nu\lambda\rho}$ is the four-dimensional Levi-Civita tensor, with $\epsilon_{0123}=1$, 
Re-expressed in terms of the dual tensors,  the constitutive equations (\ref{eq:Hmunu1}) are
\beq
^*\!H_{\mu\nu}=
\varepsilon_0\sqrt{-g}g_{\mu\lambda}g_{\nu\rho}
\,^*\!F^{\lambda\rho}.  \label{eq:constdual}
\eeq

Accordingly, we get
\begin{equation}
B_i=-\frac{\sqrt{-g}}{\varepsilon_0c^2g_{00}}g^{ij}H_j-
\frac{1}{\varepsilon_0cg_{00}} \epsilon_{ijk} g_{j0}E_k,
\end{equation}
which, in three-dimensional notation becomes
\beq
\mathbf{B} = \frac{\mu}{\varepsilon_0 c^2}\,\mathbf{H}
- \frac{\mathbf{w}}{c}\times\mathbf{E} \label{eq:const3}
\eeq
where $\mu=\varepsilon$, as defined in Eq. (\ref{eq:defepsilon}).

We notice that when $g_{\mu\nu}=\eta_{\mu\nu}$, where $\eta_{\mu\nu}$ is the Minkowski tensor, then $\mu=\varepsilon=\delta_{ij}$; for instance, this is true for spacetimes that are flat at space infinity and, in these cases, we have $\mb D=\varepsilon_{0}\mb E$, $\mb B =\mu_{0} \mb E$, where $\varepsilon_{0}\mu_{0}=c^{-2}$. 

In summary, Maxwell's equations in curved spacetime are equivalent to the flat spacetime equations (\ref{eq:maxwell}) in a medium with the constitutive equations (\ref{eq:const2}) and (\ref{eq:const3}).  The above description shows that the interaction between electromagnetic fields and the gravitational field is determined by $\varepsilon$, which describes the features of an equivalent  medium, and by the vector $\mb w$, which originates from the off-diagonal elements of the spacetime metric, which are usually related to the so-called gravitomagnetic effects  \cite{Ruggiero:2023ker}. In this framework, it is relevant to point out that both $\varepsilon$ and $\mb w$ are independent of a confarmal factor: this fact reflects the conformal invariance of Maxwell's equations \cite{masshoonprd}.

\section{The spacetime metric in the laboratory frame} \label{sec:fermi}

The spacetime metric in the laboratory frame - or proper detector frame - can be obtained using Fermi coordinates; more generally speaking, the latter  are used to express the spacetime metric in the vicinity of a reference world-line, which can be thought of as the world-line of the laboratory frame where a measurement device (the ``detector'') is placed. As showed for instance by \citet{Ni:1978di,Li:1979bz,1982NCimB..71...37F,marzlin}, \citet{Ruggiero_2020}  this expression depends both on how the laboratory frame moves in spacetime (i.e. on its acceleration and the rotation of its axes) and on the spacetime curvature through the Riemann  tensor.  Here, we are concerned with the effects produced by gravitational waves, hence we neglect the inertial effects in the definition of the metric elements: in other words we  consider a geodesic and non rotating frame. Accordingly, using  Fermi coordinates $(cT,X,Y,Z)$, up to quadratic displacements $|X^{i}|$ from the reference world-line, the line element turns out to be (see e.g.  \citet{manasse1963fermi,MTW})
\beq
ds^{2}=-\left(1+R_{0i0j}X^iX^j \right)c^{2}dT^{2}-\frac 4 3 R_{0jik}X^jX^k cdT dX^{i}+\left(\delta_{ij}-\frac{1}{3}R_{ikjl}X^kX^l \right)dX^{i}dX^{j}. \label{eq:mmmetric}
\eeq
Here  $R_{\alpha \beta \gamma \delta}=R_{\alpha \beta \gamma \delta}(T)$ is the projection of the 
Riemann curvature tensor on the orthonormal tetrad $e^{\mu}_{(\alpha)}(\tau)$ of the
reference observer, parameterized by the proper time\footnote{In $e^{\mu}_{(\alpha)}$ tetrad indices like $(\alpha)$ are within parentheses, while  $\mu$ is a  background spacetime index; however, for the sake of simplicity, we drop here and henceforth parentheses to refer to tetrad indices, which are the only ones used.} $\tau$: $\displaystyle R_{\alpha \beta \gamma \delta}(T) = R_{\alpha \beta \gamma \delta}(\tau)=R_{\mu\nu \rho
\sigma}e^\mu_{(\alpha)}(\tau)e^\nu_{(\beta)}(\tau)e^\rho_{(\gamma)}(\tau)e^\sigma 
_{(\delta)}(\tau)$ and it is evaluated along the reference geodesic, where $T=\tau$ and $ X^{i}=0$.
The line element (\ref{eq:mmmetric}) can be written in a compact form by introducing $\Phi, \mathcal A_{i}, \Psi_{ij}$ defined as
\[
\frac{\Phi}{c^{2}}=\frac{g_{00}+1}{2} \quad \frac{\Psi_{ij}}{c^{2}}=\frac{g_{ij}-\delta_{ij}}{2} \quad \frac{\mathcal A_{i}}{c^{2}}=-\frac{g_{0i}}{2},
\]
Accordingly, we get
\beq
\mathrm{d} s^2= -c^2 \left(1-2\frac{\Phi}{c^2}\right)\rmd T^2 -\frac4c \mathcal A_{i}\rmd X^{i}\rmd T  +
 \left(\delta_{ij}+2\frac{\Psi_{ij}}{c^2}\right)\rmd X^i \rmd X^j\ , \label{eq:weakfieldmetric11}
\eeq
where, explicitly:
\begin{eqnarray}
\Phi (T, { X^{i}})&=&-\frac{c^{2}}{2}R_{0i0j}(T )X^iX^j, \label{eq:defPhiG}\\
\mathcal A^{}_{i}(T ,{X^{i}})&=&\frac{c^{2}}{3}R_{0jik}(T )X^jX^k, \label{eq:defAG}\\
\Psi_{ij} (T, {X^{i}}) & = & -\frac{c^{2}}{6}R_{ikjl}(T)X^{k}X^{l}. \label{eq:defPsiG}
\end{eqnarray}
In the above definitions, $\Phi$ and $\mathcal A_{i}$ are, respectively, the \textit{gravitoelectric} and \textit{gravitomagnetic} potential, and $\Psi_{ij}$ is the perturbation of the spatial metric \cite{Mashhoon:2003ax,Ruggiero_2020,Ruggiero:2021uag,Ruggiero:2023ker}. Notice that the line element (\ref{eq:weakfieldmetric11}) is a perturbation of flat Minkowski spacetime; in other words $|\frac{\Phi}{c^{2}}| \ll 1$, $|\frac{\Psi_{ij}}{c^{2}}| \ll 1$, $|\frac{\mathcal A_{i}}{c^{2}}| \ll 1$.  

As for our purposes, we are interested in describing the field of a plane gravitational wave using this formalism. Before doing that, we briefly recall the standard approach to the description of plane gravitational waves. 
Starting from Einstein's equations
\beq
G_{\mu\nu}=\frac{8\pi G}{c^{4}}T_{\mu\nu}, \label{eq:einstein0}
\eeq
we suppose that the spacetime metric $g_{\mu\nu}$ is in the form $\displaystyle g_{\mu\nu}=\eta_{\mu\nu}+h_{\mu\nu}$, where $|h_{\mu\nu}|\ll 1$ is a small perturbation of the Minkowski tensor $\eta_{\mu\nu}$ of flat spacetime. Setting $\bar h_{\mu\nu}=h_{\mu\nu}-\frac 1 2 \eta_{\mu\nu}h$, with $h=h^{\mu}_{\mu}$, Einstein's field equations (\ref{eq:einstein0}) in the Lorentz gauge $\displaystyle \partial_{\mu} \bar h^{\mu\nu}=0$ turn out to be
\beq
\square \bar h_{\mu\nu}=-\frac{16\pi G}{c^{4}}T_{\mu\nu}, \label{eq:einstein2}
\eeq
where $\square = \partial_{\mu}\partial^{\mu}=\nabla^{2}-\frac{1}{c^{2}}\frac{\partial}{\partial t^{2}}$. The vacuum (i.e. $T_{\mu\nu}=0$) solutions of Eq. (\ref{eq:einstein2}) are gravitational waves propagating in empty space. Imposing  the transverse and traceless (TT) gauge and using the corresponding coordinates $(ct_{\mathrm{TT}},x_{\mathrm{TT}},y_{\mathrm{TT}},z_{\mathrm{TT}}$) \cite{maggiore2007gravitational,Rakhmanov_2014} the solutions for a plane wave propagating along the $x$ direction are given by 
\beq
\bar h_{\mu\nu}=-\left(h^{+}e^{+}_{\mu\nu}+h^{\times}e^{\times}_{\mu\nu} \right), \label{eq:gwsol1}
\eeq
{with
\beq
h^{+}=A^{+}\cos \left(\omega t_{\mathrm{TT}}-kx_{\mathrm{TT}} +\phi^{+} \right), \quad h^{\times}=A^{\times}\cos \left(\omega t_{\mathrm{TT}}-kx_{\mathrm{TT}} +\phi^{\times}\right), \label{eq:gwsol20}
\eeq
where $\phi^{+}, \phi^{\times}$ are constants, and the linear polarization tensors of the wave are
\beq
e^{+}_{\mu\nu}=\left[\begin{array}{cccc}0 & 0 & 0 & 0 \\0 & 0 & 0 & 0 \\0 & 0 & 1 & 0 \\0 & 0 & 0 & -1\end{array}\right], \quad e^{\times}_{\mu\nu}=\left[\begin{array}{cccc}0 & 0 & 0 & 0 \\0 & 0 & 0 & 0 \\0 & 0 & 0 & 1 \\0 & 0 & 1 & 0\end{array}\right] . \label{eq:gwsol3}
\eeq 

Here $A^{+}, A^{\times}$ are the amplitude of the wave in the two polarization states, $\phi^{+}, \phi^{\times}$ the corresponding phases,  while $\omega$ is the frequency and $k$ the wave number, so that the wave four-vector is $\displaystyle k^{\mu}=\left(\frac \omega c, k, 0, 0 \right)$, with $k^{\mu}k_{\mu}=0$.  The two linear polarizations states can be added with phase difference of $\pm \pi/2$ to get circularly polarized waves. We will use 
\beq
h^{+}=A^{+}\sin \left(\omega t_{\mathrm{TT}}-kx_{\mathrm{TT}}  \right), \quad h^{\times}=A^{\times}\cos \left(\omega t_{\mathrm{TT}}-kx_{\mathrm{TT}}\right), \label{eq:gwsol2}
\eeq
thus fixing the phase difference: accordingly, circular polarization corresponds to the condition $A^{+}=\pm A_{\times}$. In conclusion, in TT coordinates the spacetime element is given by
\beq
ds^2= -c^{2}dt_{\mathrm{TT}}^2+dx_{\mathrm{TT}}^2 +(1-h^{+})dy_{\mathrm{TT}}^2 +(1+h^{+})dz_{\mathrm{TT}}^2 -2h_{\times} dy_{\mathrm{TT}} dz_{\mathrm{TT}}\,. \label{eq:TTmetrica}
\eeq

To express the metric (\ref{eq:mmmetric}) where the curvature tensor appears,  we remember that up to linear order in the perturbation $h_{\mu\nu}$, we can write the following expressions for the Riemann tensor \cite{MTW}:
\beq
R_{ikjl}=\frac 1 2 \left(h_{il,jk}+h_{kj,li}-h_{kl,ji}-h_{ij,lk} \right) \label{eq:riemann0}
\eeq
and
\beq
R_{ij0l}=\frac 1 2 \left( h_{il,j0}-h_{jl,i0} \right). \label{eq:riemann1}
\eeq
Accordingly, we exploit the gauge invariance in linear approximation \cite{straumann2013applications}  and use the above expressions  to calculate the Riemann tensor in Fermi coordinates. 

In particular,  in the metric (\ref{eq:mmmetric}) the Riemann tensor is evaluated along the reference world-line: so, after calculating the components of Riemann tensor using Eq. (\ref{eq:gwsol2}), we set $X^{i}=0$. We point out that the expression of the metric tensor is obtained in the large wavelength limit, which means that the typical dimension $L$ of the frame  is negligible with respect to the wavelength $\lambda$; more accurate expressions can be obtained, which contains higher order terms in the small parameter $\displaystyle \epsilon = \frac{L}{\lambda}$  (see e.g. \citet{Ruggiero:2022gzl} and references therein). Actually, the series expansion can be exactly summed to obtain a compact form \cite{1982NCimB..71...37F,Berlin:2021txa}.

As a consequence, if we define the functions
\beq
f^{+}=\frac 1 2 A^{+}\sT, \quad f^{\times}=\frac 1 2 A^{\times} \cT \label{def:fpfc}
\eeq
the gravitoelectric potential (\ref{eq:defPhiG}) is written as
\beq
\Phi=\frac{\omega^{2}}{2}\left[ f^{+} Y^{2}+2f^{\times}\ YZ-f^{+} Z^{2} \right], \label{eq:defPhicomp}
\eeq

while the components of the gravitomagnetic potential (\ref{eq:defAG})  are

\begin{eqnarray}
\mathcal A_{X}&=&\frac{\omega^{2}}{3} \left[f^{+} (Y^{2}-Z^{2})+2f^{\times}\ ZY \right], \label{eq:defAX} \\
\mathcal A_{Y}&=& \frac{\omega^{2}}{3} \left[-f^{+} YX-f^{\times} ZX \right], \label{eq:defAY} \\
\mathcal A_{Z}&=& \frac{\omega^{2}}{3} \left[-f^{\times} YX+f^{+} XZ \right].  \label{eq:defAZ}
\end{eqnarray}

In addition, starting from the definition
\beq
\Psi_{ij} (T, {X^{i}})  =  -\frac{c^{2}}{6}R_{ikjl}(T)X^{k}X^{l}, \label{eq:defPsiGbis}
\eeq
we explicitly calculate the remaining metric components:
\begin{eqnarray}
\Psi_{XX} & = & -\frac 1 6 \omega^{2} \left(-\fp Y^{2}+\fp Z^{2}-2\fc YZ \right), \nonumber \\ 
\Psi_{XY} & = & -\frac 1 6 \omega^{2} \left(\fp YX+\fp ZX \right), \nonumber \\
\Psi_{XZ} & = & -\frac 1 6 \omega^{2} \left(\fc YX-\fp ZX \right), \nonumber \\
\Psi_{YY} & = & -\frac 1 6 \omega^{2} \left(-\fp X^{2} \right), \nonumber \\
 \Psi_{YZ} & = & -\frac 1 6 \omega^{2} \left(-\fc X^{2} \right), \nonumber \\
 \Psi_{ZZ} & = & -\frac 1 6 \omega^{2} \left(\fp X^{2} \right). \label{eq:defPsi}
\end{eqnarray}

\section{Modified Electromagnetic Fields} \label{sec:modfields}

We showed that it is possible to study electromagnetic fields in curved spacetime in complete analogy with the formulation of Maxwell's equations in flat spacetime in presence of an equivalent (generally non isotropic) medium. It is important to emphasize the limitations of this approach. Specifically, we assume that the electromagnetic field doesn't impact the background spacetime; in essence, we focus on how the background spacetime affects the electromagnetic fields while disregarding any reciprocal influence of these fields on the spacetime itself.  Of course, this is not the general case, in fact in extreme astrophysical events the electromagnetic field can be source of the gravitational field, and the coupled Maxwell's and Einstein's equations must be solved. However, this approach is adequate for our purposes. 

In what follows, we suppose to have and electric field or a magnetic field before the passage of the wave: we are interested in understanding how these fields are perturbed by the passage of the wave. As we discussed above, the natural framework to study  this interaction is the laboratory metric (\ref{eq:weakfieldmetric11}), with the definitions (\ref{def:fpfc})-(\ref{eq:defPsi}): {in other words, we consider a reference frame which, before the passage of the gravitational wave, is endowed by a flat space-time metric (see Eq.(\ref{eq:weakfieldmetric11}) when the Riemann tensor is zero). In this frame, the unperturbed electric and magnetic fields are solution of Maxwell's equations (see also the discussion in \citet{Ratzinger:2024spd}). }

Let us start by the case when, before the passage of the gravitational wave, an electric field $\mb E^{0}$ is present. Hence, we consider  the first of Eqs. (\ref{eq:maxwell}) with the constitutive equation (\ref{eq:const1}); notice that a magnetic field eventually produced by the gravitational wave, does not enter this equation, since both $w_{i}$ and $H_{i}$ are first order terms in the wave perturbation, so if we work at linear order their contribution can be neglected. As a consequence, the constitutive equation reads
\beq
D^{j}=-\epsilon_{0} \frac{\sqrt{-g}}{g_{00}}g^{ji}E_{i}. \label{eq:const11}
\eeq

We work at linear order with respect to the wave amplitude, so we may write
\beq
-\frac{\sqrt{-g}}{g_{00}} \simeq 1+f, \label{eq:pert1}
\eeq
\beq
g^{ji} \simeq \delta^{ji}+f^{ji}, \label{eq:pert2}
\eeq
\beq
\sqrt{-g} \simeq 1+ p, \label{eq:pert22}
\eeq
where  $f$, $f^{ji}$ and $p$ are proportional to the wave amplitude, i.e. they are linear functions of $A^{+}, A^{\times}$. In addition, we look for a solution for the electric field in the form  $E_{i}=E^{0}_{i}+e_{i}$ where, again, $e_{i}$ is a linear function of   $A^{+}, A^{\times}$. So, up to first order in the wave amplitude, we may write Eq. (\ref{eq:const11}) in the form
\beq
D^{j}=\left[\delta^{ij}\left(1+f \right)+f^{ji} \right]E_{i}^{0}+\delta^{ji} e_{i}. \label{eq:pert3}
\eeq
As a consequence:
\beq
\partial_{j}D^{j}= \partial_{j}\left(\delta^{ji}E_{i}^{0} \right)+\partial_{j} \left[\left(\delta^{ji}f+f^{ji}\right) E^{0}_{i} \right]+\partial_{j}\left(\delta^{ji}e_{i} \right). \label{eq:pert4}
\eeq
According to Eq. (\ref{eq:sources}), we have  $\rho=(1+p)\rho_{0}$ where $\rho_{0}$ is the ``true'' charge density, i.e. the source of the unperturbed electric field $\mb E^{0}$:
\beq
\partial_{j}E^{0\, j}=\frac{\rho_{0}}{\epsilon_{0}}. \label{eq:pert6}
\eeq
Then, the  first of Eqs.  (\ref{eq:maxwell}) can be written as
\beq
\partial_{j}\left(\delta^{ji}E_{i}^{0} \right)+\partial_{j} \left[\left(\delta^{ji}f+f^{ji}\right) E^{0}_{i} \right]+\partial_{j}\left(\delta^{ji}e_{i} \right)=\frac{(1+p)\rho_{0}}{\epsilon_{0}}. \label{eq:pert5}
\eeq
Taking into account Eq. (\ref{eq:pert6}), we finally have
\beq
\partial_{j}\left(\delta^{ji}e_{i} \right)=\frac{p\rho_{0}}{\epsilon_{0}}- \partial_{j} \left[\left(\delta^{ji}f+f^{ji}\right) E^{0}_{i} \right]. \label{eq:pert7}
\eeq
We see that the effective charge density
\beq
\displaystyle \frac{\rho_{eff}}{\epsilon_{0}}=\frac{p\rho_{0}}{\epsilon_{0}}- \partial_{j} \left[\left(\delta^{ji}f+f^{ji}\right) E^{0}_{i} \right] \label{eq:rhoeff}
\eeq
becomes source for the perturbation of the electric field
\beq
\partial_{i}e^{i} =\frac{\rho_{eff}}{\epsilon_{0}}.  \label{eq:pert8}
\eeq
From Eq. (\ref{eq:rhoeff}), we observe that in vacuum, where $\rho_{0}=0$, gravitational waves interact with the unperturbed electric field, determining  an effective charge density. This leads to a perturbed electric field similar to that found in a medium rather than in vacuum.

Now, let us consider the case of the presence of a magnetic field $\mb B^{0}$ before the passage of the wave. To this end, we consider the second of the Maxwell's equations (\ref{eq:maxwell})  with the constitutive equation (\ref{eq:const3}): exactly as before, an electric field eventually produced by the gravitational wave does not enter this constitutive equation, since both $w_{i}$ and $E_{i}$ are first order terms in the wave perturbation.

From Eq. (\ref{eq:const3}) we may write
\beq
H_{i}=-\epsilon_{0}c^{2}\frac{g_{00}}{\sqrt{-g}} \gamma_{ij} B^{j}, \label{eq:2pert}
\eeq
where $\gamma_{ij}=g_{ij}-\frac{g_{i0}jg_{j0}}{g_{00}}$ is the spatial metric \cite{Ruggiero:2023ker};  at linear order we have $\gamma_{ij}=g_{ij}$. As before, we write
\beq
-\frac{g_{00}}{\sqrt{-g}}=1+\ell  \label{eq:2pert1}
\eeq
where $\ell=-f$,
\beq
g_{ij}=\delta_{ij}+s_{ij},  \label{eq:2pert2}
\eeq
\beq
B_{j}=B^{0}_{j}+b_{j}. \label{eq:2pert3}
\eeq
Again, we observe that $\ell, s_{ij}$ and $b_{j}$ are linear functions of the gravitational wave amplitudes $A^{+},A^{\times}$.

Accordingly, at linear order in the perturbation due to the gravitational wave, we may write the lhs member of the second Maxwell's equation (\ref{eq:maxwell}) $\epsilon_{kmi} \partial_{m} H_{i}$  in the form
\beq
\epsilon_{kmi}\partial_{m} \left[ \left(\delta_{ij}+s_{ij}+\delta_{ij}f \right)B^{0}_{j}+\delta_{ij} b_{j} \right]. \label{eq:2pert4}
\eeq
Then,  the source equation for the magnetic field reads
\beq
\epsilon_{kmi}\partial_{m} \delta_{ij}B^{0}_{j}+\epsilon_{kmi} \partial_{m}\left[\left(s_{ij}+\delta_{ij}f  \right)B^{0}_{j} \right]+\epsilon_{kmi}\partial_{m}\left(\delta_{ij} b_{j}\right)=\mu_{0}\left(1+p \right)j^{0}_{k}. \label{eq:2pert5}
\eeq
where $j_{k}^{0}$ is the ``true'' current which appears in the equation for the unperturbed magnetic field
\beq
\epsilon_{kmi} \partial_{m} B^{0}_{i}=\mu_{0} j^{0}_{k}. \label{eq:2pert0}
\eeq
Consequently, we may write Eq. (\ref{eq:2pert5}) in the form
\beq
\epsilon_{kmi}\partial_{m}\left(\delta_{ij} b_{j}\right)=\mu_{0} p j^{0}_{k}-\epsilon_{kmi} \partial_{m}\left[\left(s_{ij}+\delta_{ij}f  \right)B^{0}_{j} \right]. \label{eq:2pert6}
\eeq

If we define the effective current
\beq
\mu_{0} j_{k}^{eff}=\mu_{0} p j^{0}_{k}-\epsilon_{kmi} \partial_{m}\left[\left(s_{ij}+\delta_{ij}f  \right)B^{0}_{j} \right] \label{eq:effcurr11}
\eeq
 we finally may write
\beq
\epsilon_{kmi}\partial_{m}b_{i}=\displaystyle \mu_{0} j_{k}^{eff}. \label{eq:2pert7}
\eeq
Once more, we observe that the effective current, originating from the interaction between the gravitational field and the initial magnetic field, acts as a source for the perturbation $b_{i}$.

In summary, at linear order  in the gravitational wave amplitude, we have obtained the source equations  for the perturbed electric (\ref{eq:pert8}) and magnetic (\ref{eq:2pert7}) field.

\section{Some Examples} \label{sec:examples}

Below, we will apply the formalism described so far to solve Maxwell's equations in some particular cases. We point out that Eqs. (\ref{eq:pert8}) and (\ref{eq:2pert7}) correspond to the perturbed Maxwell's equations for the \textit{sources} of the electric and magnetic field. However, in order to fulfill the whole set of Maxwell's equations, we need to consider also the homogenous ones. To this end, as it customary, we will introduce a scalar and vector potential.

\subsection{A Uniform Electric Field} \label{sec:E0}

We suppose that before the passage of the wave we have a uniform electric field $\mb E^{0}=E^{0}\mb u_{Z}$, where $E^{0}=\mathrm{const}$. To fix ideas,  we can consider the field within a capacitor.

We need to solve Eq. (\ref{eq:pert8}), with $\rho_{0}=0$. In order to solve also the homogenous Maxwell equations, we set $\mb e=-\nabla \phi$. Before the passage of the wave, we have $\displaystyle \phi_{0}(Z)=V_{0}\left(1-\frac{Z}{d}\right)$, where $V_{0}$ is the difference of potential  between the two armatures, and $d$ the distance between them. In this case, $E^{0}=V_{0}/d$ is the unperturbed constant electric field, and $\phi_{0}$ satisfies the Laplace equation. 

Accordingly,  Eq. (\ref{eq:pert8}) becomes
\beq
\nabla^{2}\phi=-\frac{\rho_{eff}}{\epsilon_{0}}. \label{eq:ex1}
\eeq
Taking into account the expression of the line element (\ref{eq:weakfieldmetric11}), with the definitions (\ref{def:fpfc})-(\ref{eq:defPsi}), we obtain the following effective charge density:
\beq
\rho_{eff}=-\epsilon_{0} \frac{\omega^{2}}{c^{2}} \frac{5    \left(A^{\times} \cos \! \left(\omega  T \right) Y -A^{+} \sin \! \left(\omega  T \right) Z \right) \mathit{E^{0}}}{6}.  \label{eq:ex2}
\eeq

The solution of Eq. (\ref{eq:ex1}) with the conditions $\phi(Y,0)=V_{0}=E^{0}d, \phi(Y,d)=0$ turns out to be:
\beq
\phi(Y,Z)=E^{0}(d-Z)\left[1+\frac{5}{36}\frac{\omega^{2}}{c^{2}}A^{+}  \sin \! \left(\omega  T \right) \left(d+Z \right)Z-\frac{5}{12}\frac{\omega^{2}}{c^{2}}A^{\times}  \sin \! \left(\omega  T \right)YZ   \right]. \label{eq:ex8}
\eeq

Accordingly,  the electric field is in the form $\mb E= E_{Y}\mb u_{Y}+E_{Z} \mb u_{Z}$, where
\beq
E_{Y}=\frac{5}{12}E^{0}(d-Z)Z \frac{\omega^{2}}{c^{2}}A^{\times} \cos \! \left(\omega  T \right),  \label{eq:Eyfin}
\eeq

\beq
E_{Z}=E^{0}\left[1+\frac{5}{12}(d-2Z)Y \frac{\omega^{2}}{c^{2}} A^{\times}\cos \! \left(\omega  T \right)-\frac{5}{36}\left(d^{2}-3Z^{2} \right)\frac{\omega^{2}}{c^{2}}A^{+} \sin \! \left(\omega  T \right)   \right] ,\label{eq:Ezfin}
\eeq

\begin{figure}[t]
\begin{center}
\includegraphics[scale=.3]{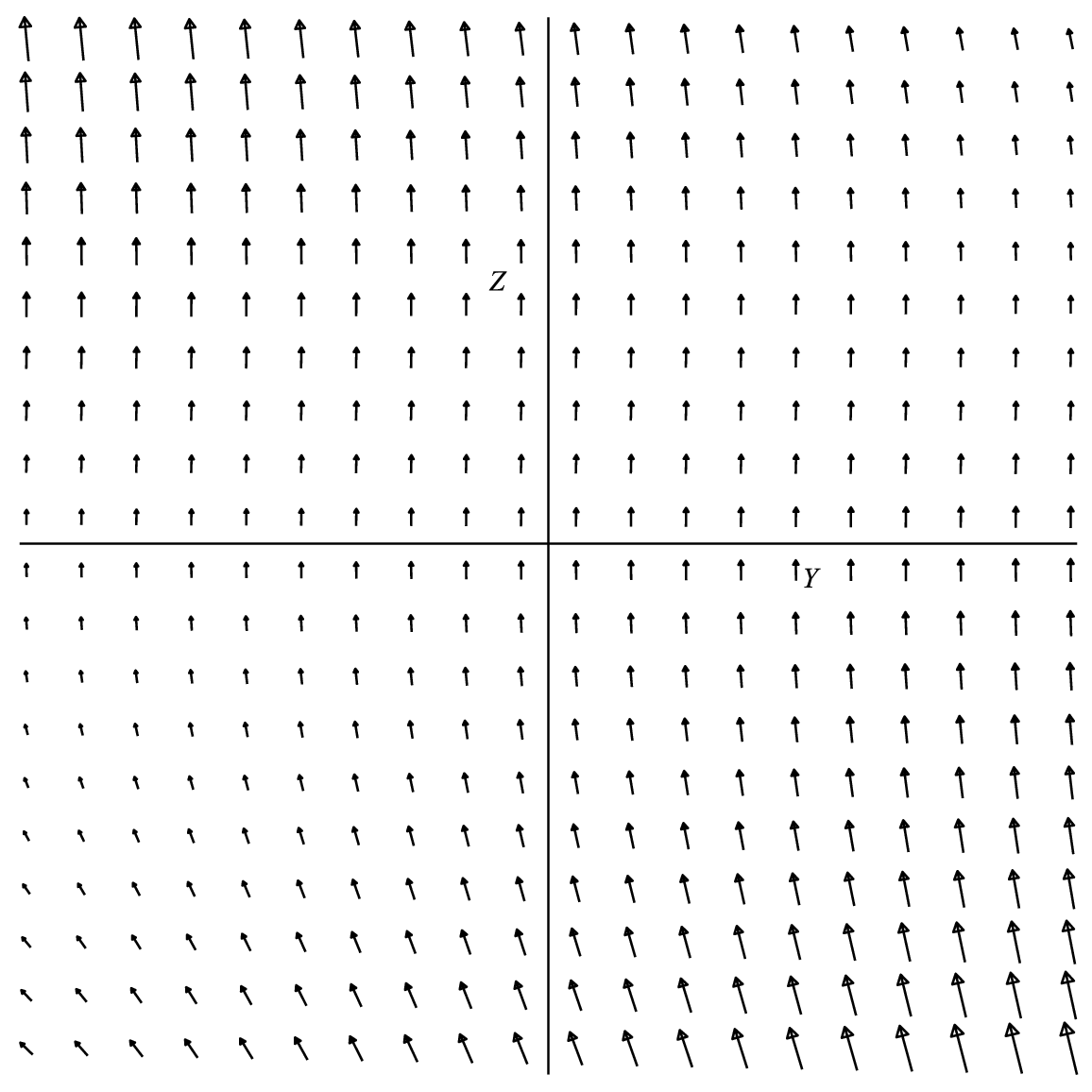} \hspace{1cm}
\includegraphics[scale=.3]{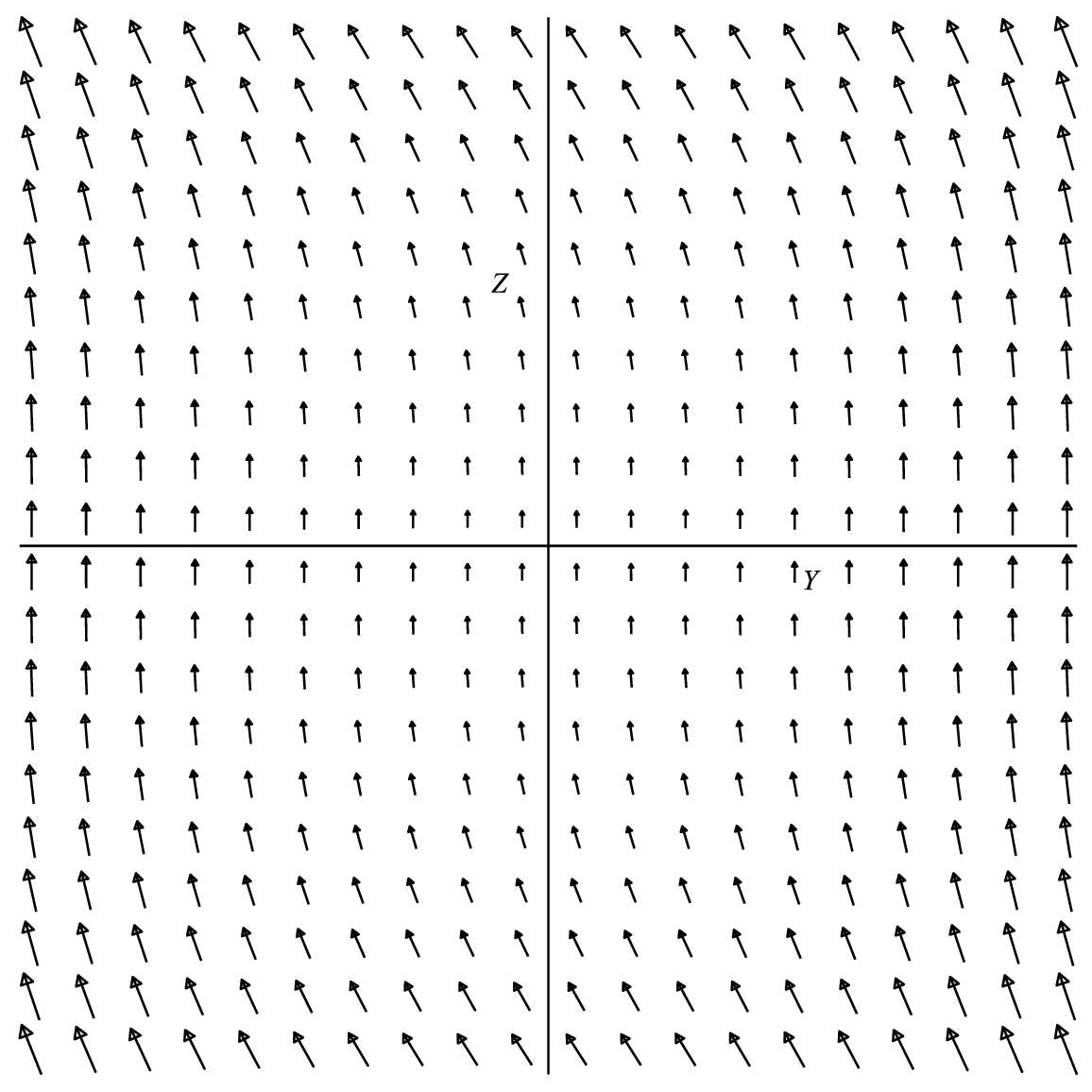}
\caption{On the left: the qualitative  effect of the gravitational wave on the (initially uniform) electric field $\mb E$ in the plane $YZ$; on the right, the same for the magnetic field $\mb B$. Both figure shows the field at a fixed time, such that $\omega T=\pi/3$.  The effects of the wave perturbation are greatly increased to make them visibile: see the discussion in Section \ref{sec:discon}. } \label{fig:campoEB}
\end{center}
\end{figure}

A sketch of his electric field is plotted in Figure \ref{fig:campoEB}.

\subsection{A Uniform Magnetic Field} \label{sec:B0}

Let us consider the case when a uniform magnetic field $\mb B^{0}=B^{0} \mb u_{Z}$ is present before the passage of the wave. To this end, we need to solve Eq. (\ref{eq:2pert7}) and, in addition, in order to solve also the homogenous Maxwell's equation for the magnetic field, we set $\mb b= \nabla \times \mb A$, where $\mb A$ is the vector potential. Accordingly, exploiting the gauge freedom and imposing the Coulomb gauge ($\bm \nabla \cdot \mb A=0$), we may write Eq.   (\ref{eq:2pert7})  in the form
\beq
\nabla^{2} \mb A=-\mu_{0} \mb j^{eff}, \label{eq:2expoisson}
\eeq 
where, in this case, from Eqs. (\ref{eq:weakfieldmetric11}), (\ref{def:fpfc})-(\ref{eq:defPsi}), we obtain
\begin{eqnarray}
j^{eff}_{X} & = & \frac{\omega^{2}}{c^{2}}\frac{\left(2 A^{+} \sin \! \left(\omega  T \right) Y +2 A^{\times} \cos \! \left(\omega  T \right) Z \right) }{3} B^{0}, \nonumber \\
j^{eff}_{Y} & = & - \frac{\omega^{2}}{c^{2}}\frac{ A^{+} \sin \! \left(\omega  T \right) X }{2} B^{0}, \label{eq:jmod} \\
j^{eff}_{Z} & = & -\frac{\omega^{2}}{c^{2}}\frac{ A^{\times} \cos \! \left(\omega  T \right) X }{2} B^{0}. \nonumber
\end{eqnarray}
On solving Eq. (\ref{eq:2expoisson}), taking into account the vacuum unperturbed solution $\mb A^{0}$ such that $\mb B^{0}=\bm \nabla \times \mb A^{0}$ where
\beq
A^{0}_{X}=-\frac{B^{0}Y}{2}, \quad A^{0}_{Y}=\frac{B^{0}X}{2}, \quad A^{0}_{Z}=0 \label{eq:A0},
\eeq
we have
\begin{eqnarray}
A_{X} & = & -\frac{\omega^{2}}{c^{2}}\frac{ A^{+} \sin \! \left(\omega  T \right)  \,Y^{3}}{9}B^{0}-\frac{\omega^{2}}{c^{2}}\frac{  A^{\times} \cos \! \left(\omega  T \right) Z^{3}}{9}B^{0}-\frac{Y \mathit{B^{0}}}{2}, \label{eq:Axsol} \\
A_{Y} & = & \frac{\omega^{2}}{c^{2}}\frac{  A^{+} \sin \! \left(\omega  T \right) X  \,Z^{2}}{4}B^{0}+\frac{X \mathit{B^{0}}}{2},  \label{eq:Aysol} \\
A_{Z} & = &  \frac{\omega^{2}}{c^{2}}\ \frac{  A^{\times} \cos \! \left(\omega  T \right) X  \,Z^{2}}{4}B^{0},  \label{eq:Azsol} 
\end{eqnarray}
and, consequenty the magnetic field $\mb B$ has the following components
\begin{eqnarray}
B_{X} & = & -\frac{\omega^{2}}{c^{2}}\frac{  A^{+} \sin \! \left(\omega  T \right) XZ}{2} B^{0}  \label{eq:Bxsol} \\
B_{Y} & = &  -7\frac{\omega^{2}}{c^{2}}\frac{  A^{\times} \cos \! \left(\omega  T \right) Z^{2}}{12}B^{0} \label{eq:Bysol} \\
B_{Z} & = &  B^{0}\left[1+\frac{\omega^{2}}{c^{2}}\frac{  A^{+} \sin \! \left(\omega  T \right)  \,Z^{2}}{4} +\frac{\omega^{2}}{c^{2}}\frac{ A^{+} \sin \! \left(\omega  T \right)  \,Y^{2}}{3}\right]  \label{eq:Bzsol} 
\end{eqnarray}

This magnetic field in the plane $YZ$ is plotted in Figure \ref{fig:campoEB}.

\section{Discussion and Conclusions} \label{sec:discon}

We studied the interaction of a plane gravitational wave with electromagnetic fields.  In particular, we  focused on the perturbation induced by a gravitational wave on  electromagnetic fields that are present before its passage. Working at first order in the wave amplitude, we showed that the perturbed electromagnetic fields satisfy equations Eqs.  (\ref{eq:pert8}) and (\ref{eq:2pert7}), where new terms are present which originate  from the coupling of the gravitational field of the wave with the sources of the unperturbed fields and with the unperturbed fields themselves. {This coupling results in an  effective charge density and current, which are the sources of the perturbed fields: this fact was already identified in previous works  \citep{Berlin:2021txa,PRLcav,Ratzinger:2024spd}, where both the TT frame and the laboratory frame are considered. 
{Our approach is novel as it combines the formulation of Maxwell's equations in curved spacetime as flat spacetime equations in an effective medium with the explicit use of the proper detector frame. By employing Fermi coordinates, this framework allows direct reference to measurable quantities.}  {Naturally, measurable effects are independent of the method used to parameterize spacetime events; however, we believe that the laboratory frame approach may be preferable due to its descriptive simplicity.} Given the world-line of the laboratory, the Fermi metric allows the measurement process to be described in a coherent way from a relativistic point of view: generally, this metric depends on both  the motion of the laboratory in a background spacetime and  the spacetime curvature itself. For simplicity, we neglected the actual motion of a terrestrial laboratory which is not in free fall, since we are interested only in the time-depending effects produced by a monochromatic gravitational wave. In this approach, the unperturbed fields are assumed to exist and be measurable in the laboratory frame before the passage of the wave.}

To give an idea of the impact of gravitational waves on electromagnetic fields, we considered two simple situations:  the case where static and uniform electric and magnetic fields are present. These situations ideally correspond to the field within a capacitor and a within a solenoid.  In both cases (see Eqs. (\ref{eq:Eyfin})-(\ref{eq:Ezfin}) and (\ref{eq:Bxsol})-(\ref{eq:Bzsol})) we showed not only that the preexisting fields are perturbed along their initial direction, but also that components in new directions arise. In particular, the perturbations  are in the order of
\beq
\frac{\Delta E}{E^{0}} \simeq h \frac{L^{2}}{\lambda^{2}}, \quad \frac{\Delta B}{B^{0}} \simeq h \frac{L^{2}}{\lambda^{2}}, \label{eq:stime}
\eeq
where $h \simeq A^{+} \simeq A^{\times}$ is the wave amplitude, $\displaystyle \lambda \simeq \left(\frac{\omega}{c}\right)^{-1}$ is the wavelength and $L$ is a typical dimension of the measurement device.  

We remark that we used a reference frame adapted to the direction of propagation of the wave: namely the $X$ axis is the direction of propagation, and $YZ$ is the plane orthogonal to it. In general, the laboratory frame has an arbitrary orientation with respect to the direction of propagation of the wave so we should introduce a set of coordinates that are adapted to the laboratory: this is usually accomplished by using the  Euler angles (see e.g. \cite{tintoschutz}). However, this would not change the order of magnitude and the physical meaning of the effects that we are studying.  

What is the physical interpretation of the modified electric (\ref{eq:Eyfin})-(\ref{eq:Ezfin}) and magnetic  (\ref{eq:Bxsol})-(\ref{eq:Bzsol}) fields? We must remember that gravitational waves produce tidal effects, as it is manifest since their impact in the laboratory metric (\ref{eq:mmmetric}) enters through the components of the Riemann tensor. As a consequence, these effects cannot be detected locally, e.g. along the reference world-line, but only by means of a comparison of what happens at different locations in the laboratory frame. To fix ideas, if we consider two identical capacitors, one at the origin of our frame, and the other at the location $(X=X_{0},Y=Y_{0},Z=Z_{0})$ what we expect is a relative variation of the electric field as described by (\ref{eq:Eyfin})-(\ref{eq:Ezfin}); the same happens for the magnetic field. As a matter of fact,  the passage  of gravitational waves also in presence of \textit{static} electric and magnetic field produce \textit{time-varying} electromagnetic fields with the same frequency of the waves.

{In our view, the tidal nature of the perturbation induced by a gravitational wave was not clearly emphasised in previous works \citep{Berlin:2021txa,PRLcav,Ratzinger:2024spd}, possibly because their focus  was on studying the effect on specific devices such as cavities (see below). Additionally, our approach highlights  another significant consequence. In fact, as we discussed elsewhere \citep{Ruggiero_2020}, the effect of the passage of gravitational waves on test masses can be described in terms of a Lorentz-like force equation where the tidal gravitoelectromagnetic fields $\mb E^{G},\mb B^{G}$ appear. These fields can be derived from the gravitoelectric (\ref{eq:defPhicomp}) and gravitomagnetic (\ref{eq:defAX})-(\ref{eq:defAZ}) potentials and, in the frame considered, they have the following components
\small
\beq
E^{G}_{X}  = 0, \quad E^{G}_{Y}  = -\frac{\omega^{2}}{2}\left[A^{+} \sin \left(\omega T \right)Y+A^{\times} \cos \left(\omega T \right) Z \right], \quad E^{G}_{Z}  = -\frac{\omega^{2}}{2}\left[A^{\times}\cT Y-A^{+}\sT Z \right], \label{eq:defExyz1}
\eeq
\beq
B^{G}_{X}  = 0, \quad B^{G}_{Y}  = -\frac{\omega^{2}}{2}\left[-A^{\times} \cT Y+A^{+} \sT Z \right], \quad B^{G}_{Z}  = -\frac{\omega^{2}}{2}\left[A^{+}\sT Y+A^{\times}\cT Z \right]. \label{eq:defBxyz1}
\eeq
\normalsize
By comparison with the perturbed  electric (\ref{eq:Eyfin})-(\ref{eq:Ezfin}) and magnetic  (\ref{eq:Bxsol})-(\ref{eq:Bzsol}) fields, it is manifest that the tidal electromagnetic perturbations are \textit{quadratic} in the distance parameter with respect to the reference world-line, while the purely gravitational ones (\ref{eq:defExyz1}), (\ref{eq:defBxyz1}) are \textit{linear}. {The different scaling of the electromagnetic and purely gravitational perturbations can be also inferred by comparing the electromagnetic potentials (\ref{eq:ex8}), (\ref{eq:Axsol}), (\ref{eq:Aysol}), (\ref{eq:Azsol}) with the gravitational ones (\ref{eq:defPhicomp}), (\ref{eq:defAX}), (\ref{eq:defAY}), (\ref{eq:defAZ}).} {Consequently, the effects caused by electromagnetic perturbations are expected to be more challenging to detect. Currently, interferometers measure gravitoelectric effects described by (\ref{eq:defExyz1}), while gravitomagnetic effects act on test particles that are already in motion before the passage of the wave or on spinning particles \cite{Ruggiero_2020}. }

{As a consequence of the modification of the electromagnetic fields in the examples we focused on, the fields are no longer uniform during the passage of the wave, since they now depend on spatial location. This, in turn, can lead to new, peculiar effects. Consider, for instance, an electric dipole $\mb p$: before the wave passes, a uniform electric field exerts no force on the dipole. However, the perturbed field is no longer uniform, so a time-depending force $\mb F = (\mb p \cdot \nabla) \mb E$ arises. To calculate this  force, we must account for the fact that the test particles constituting the dipole are also displaced by the gravitational wave; however, at first order in the wave perturbation, this effect can be neglected. Consequently, a rough estimate of the force  magnitude is given by $\displaystyle F \simeq p E_{0} \frac{\omega^{2}}{c^{2}} h d$, where $d$ is the separation between the two dipole charges. While these effects are very small, as we discuss below, they could potentially be observed as a change in the polarization of a macroscopic sample. We have previously proposed that gravitational waves can be detected by observing modifications in electromagnetic properties, specifically referencing changes in magnetization \cite{Ruggiero_2020}. In future work, we aim to develop a comprehensive approach to exploring the effects of gravitational waves on macroscopic electromagnetic properties. }

The electromagnetic perturbations determined by the passage of a gravitational wave have the order of magnitude expressed by Eq. (\ref{eq:stime}).  To estimate them it is important to remember that  the laboratory metric (\ref{eq:mmmetric}) is approximated up to second order in $\displaystyle \frac{L}{\lambda}$.  That being said, and considering that the typical amplitude of gravitational waves reaching the Earth is $h \sim 10^{-21}$, we see that relative magnitude of the perturbations can be of the order of $\sim 10^{-23}$.  Notice that previous estimates such as those reported by  \citet{lobo}  overestimated the impact of the effect, since they are developed in the TT frame where, however, quantities have not a direct meaning in terms of observables.

 In any case, as discussed by \citet{lobo} (see also references therein), one possibility to detect very small magnetic fields is the use of SQUIDS which can measure magnetic field ranging from $10^{-15}$ T to $10^{-18}$ T \cite{carugno}; however, the order of magnitude of the effects we are considering suggests that even using such devices huge unperturbed magnetic field are needed, which are really out of our current technological possibilities. Similar considerations apply to electric fields.

Another kind of approach is the use of resonant cavities (see e.g. \citet{Berlin:2021txa}): the basic idea is that the passage of a gravitational wave affects the shape of a cavity and alters the magnetic flux passing through it, resulting in an electric signal that can be measured. In addition, as we showed, the passage of the wave produces an effective current in Maxwell’s equations, causing an electromagnetic field to oscillate at the same frequency as the gravitational wave. Recent proposals of experimental setups seem quite promising for the detection of gravitational waves \cite{Schmieden:2023fzn}. 

The general feeling is that, even if current technology does not allow these effects to be measured in terrestrial laboratories, this possibility is not excluded in the future:  {to this end, we believe that the descriptive simplicity of the laboratory frame approach is important for gaining better physical insight, as demonstrated by the examples we have considered.}

 A completely different scenario is offered by astrophysical events, where the combination of enormous electromagnetic fields and more intense gravitational waves can open up new interesting possibilities; however, this would require a different analysis than the one used in this work as it requires the simultaneous solution of Einstein's and Maxwell's equations.

In conclusion, our approach lends itself to a very simple description of the interaction of electromagnetic devices with gravitational waves in terms of observable quantities, and can be used for further studies in this field;  the possibility of developing a real experimental proposal is in any case guided by technological developments.

\section*{Acknowledgements}

The author expresses gratitude to Dr. Antonello Ortolan for engaging discussions and valuable insights. Special appreciation is extended to the local research project "Modelli gravitazionali per lo studio dell'universo" (2022) from the Department of Mathematics "G. Peano," Università degli Studi di Torino, as well as to the Gruppo Nazionale per la Fisica Matematica (GNFM) for their contributions.

\bibliography{GEM_GW}
\end{document}